\RequirePackage{lineno} 
\documentclass{emulateapj}
\modulolinenumbers[5]

\newcommand{\lat}{{\em Fermi}-LAT}

\newcommand{\far}{{\it false-association rate}}
\newcommand{\rec}{{\it recall}}

\usepackage{amssymb}
\usepackage{amsmath}
\usepackage{url}
\usepackage{natbib}
\mathchardef\mhyphen="2D

\begin{document}
\title{Search for gamma-ray-emitting active galactic nuclei in the \emph{Fermi}-LAT  
unassociated sample using machine learning}

\author{M.~Doert\altaffilmark{1,3}, M. Errando\altaffilmark{2}}
\altaffiltext{1}{Fakult\"{a}t Physik, Technische Universit\"{a}t Dortmund, 44221 Dortmund, Germany; 
\mbox{marlene.doert@tu-dortmund.de}}
\altaffiltext{2}{Department of Physics and Astronomy, Barnard College, Columbia University, NY 
10027, USA; \mbox{errando@astro.columbia.edu}}
\altaffiltext{3}{now at Department of Physics, Columbia University, New York, NY 10027, USA}

\begin{abstract}
\noindent
The second \lat\ source catalog (2FGL) is the deepest all-sky survey available in the gamma-ray band. It 
contains 1873 sources, of which 576 remain unassociated. Machine-learning algorithms can be trained on the 
gamma-ray properties of known active galactic nuclei (AGN) to find objects with 
AGN-like properties in the unassociated sample. This analysis finds 231  
high-confidence AGN candidates, with increased robustness provided by intersecting two complementary 
algorithms. A method to estimate the performance of the 
classification algorithm is also presented, that takes into account the differences between 
associated 
and unassociated gamma-ray sources. Follow-up observations targeting AGN candidates, or studies of 
multiwavelength archival data, will reduce the number of unassociated gamma-ray sources and contribute to a 
more complete characterization of the population of gamma-ray emitting AGN.
\end{abstract}


\maketitle

\section{Introduction}
The identification of astrophysical MeV and GeV sources has been a long-standing question in 
gamma-ray astronomy, mainly due to the 
limited angular resolution of gamma-ray detectors. 
The pioneer \emph{SAS-2} and \emph{COS-B} satellites 
reported 
detections of 26 sources with median location error of $\sim 1^{\circ}$ \citep{sas2,cosb}. However, 
only the emission from the \object{Crab} and \object{Vela} pulsars and the quasar \object[3C 273]{3C~273} could be firmly identified.
The deeper survey by EGRET
reported 271 gamma-ray 
sources with a median location error of $0.65^{\circ}$, but  only 101 identifications were reported 
\citep{3eg}. 

The Large Area Telescope (LAT) aboard the \emph{Fermi Gamma-ray Space Telescope} started operations 
in 2008. The increased effective area, reduced dead time, and use of silicon tracker technology 
resulted in an order of magnitude improvement in source location compared to its predecessors. The 
second \lat\  source catalog \citep[2FGL, ][]{2fgl} characterizes 1873 gamma-ray sources between 
0.1 and 100\,GeV with a median location error of $0.07^{\circ}$. 
A total of 1297 sources in the 2FGL are either identified through variability or morphology, or 
reliably associated with counterparts from  
catalogs of candidate gamma-ray-emitting source classes.
The remaining 576 sources for which no 
counterpart was identified are left unassociated. 

\begin{table} [b]
\centering
	\begin{tabular}{cccc}
		\hline\hline
		Class & Description & Source count& Label \\
		\hline
		bzb & BL Lac-type blazar& 436 & AGN \\
		bzq & FSRQ-type blazar & 370 & AGN \\
		agu & AGN of uncertain type & 257 & AGN \\
		agn & Non-blazar AGN & 11 & AGN \\
		rdg & Radio galaxy & 12 & AGN \\
		sey & Seyfert galaxy & 6 & AGN \\
		psr & Pulsar& 108 & non-AGN \\
		glc & Globular cluster& 11 & non-AGN \\
		snr & Supernova remnant & 10 & non-AGN \\
		pwn & Pulsar wind nebula& 3 &  non-AGN \\
		spp & SNR / PWN & 58 &  non-AGN \\
		hmb & High-mass binary& 4 & non-AGN \\
		nov & Nova & 1 & non-AGN \\
		gal & Normal galaxy & 6 & non-AGN \\
		sbg & Starburst galaxy& 4 & non-AGN \\
		 & Unassociated sources& 576 & \\
		\hline\hline
\end{tabular}
\caption{List of source classes 
in the 2FGL catalog.}
\label{tab:classes}
\end{table}
\begin{figure*}[thpb]
 \center
 \includegraphics[width=0.99\textwidth]{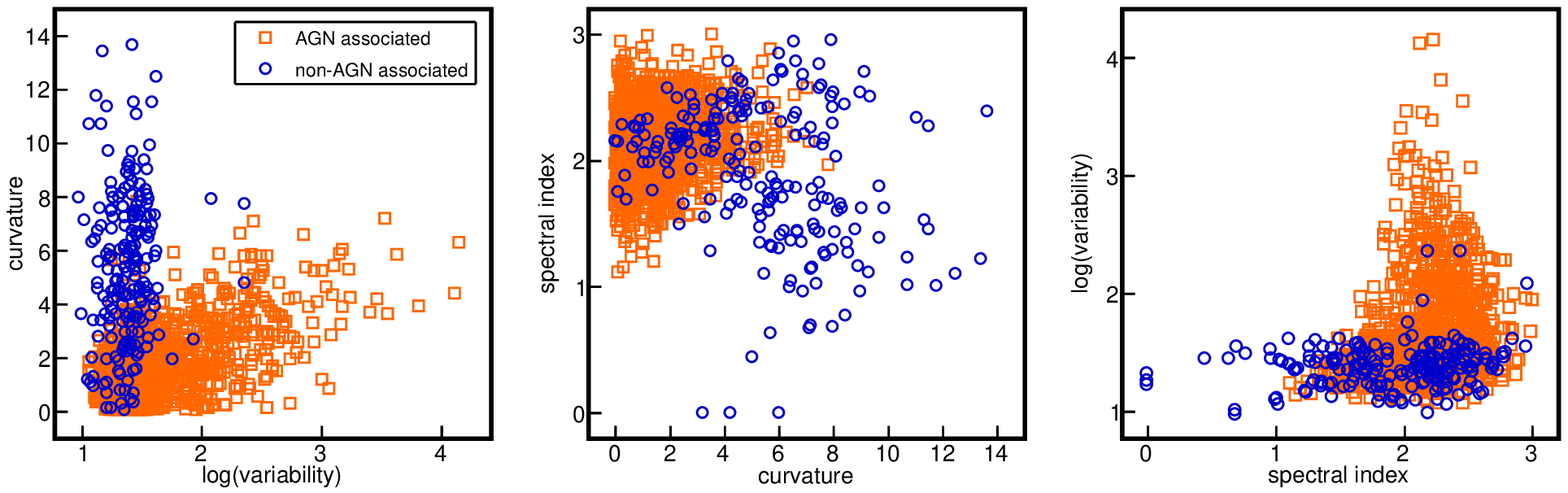} 
 \includegraphics[width=0.99\textwidth]{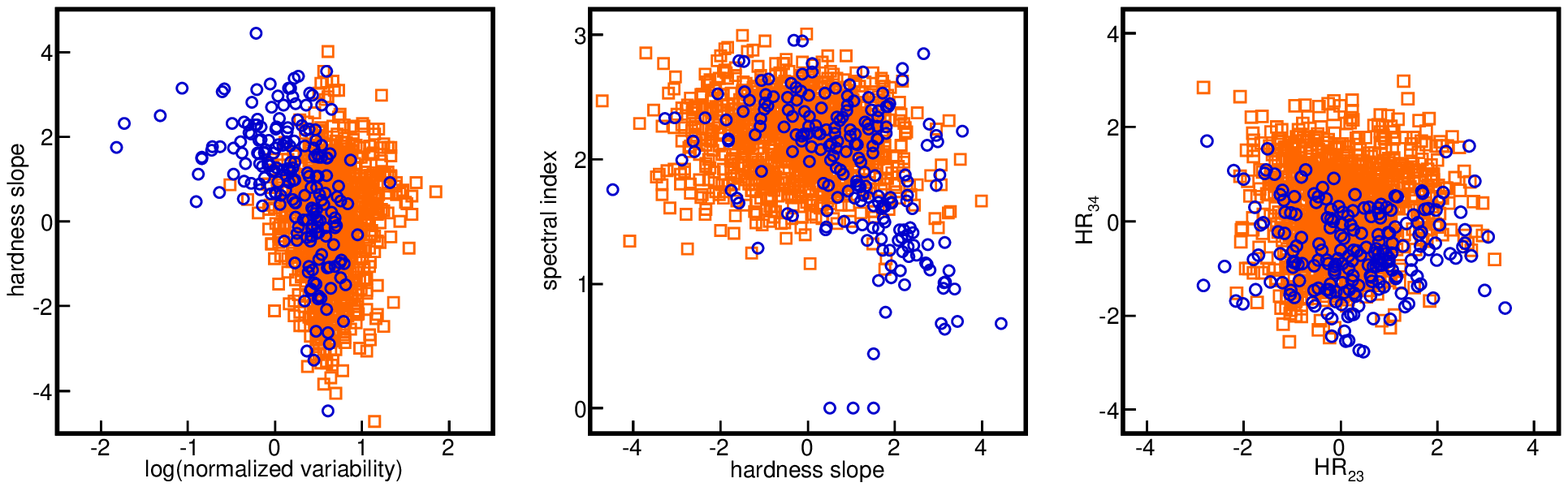} 
 \caption{Scatter plots showing some gamma-ray properties listed in the 2FGL catalog ({\it top 
panels}) and parameters used by the machine-learning algorithms ({\it bottom 
panels}, defined in Section~\ref{sec:prepare}) for AGN and non-AGN sources. 
\label{fig:2fgl_par}}
 \end{figure*}

Several studies have addressed the nature of unidentified gamma-ray sources. Searches for 
counterparts have been carried out through dedicated observations of the source error boxes \citep[e.g., 
][]{ozel,crawford} or cross-correlations with multiwavelength catalogs 
\citep{romero,sowards-emmerd}. Others studied latitude and flux distributions to 
discriminate between different source populations in the unidentified sample \citep{lamb, gehrels}, 
or compared the properties of unidentified sources to those of candidate source populations 
\citep{mukherjee,kaaret,merck}. 
A new possibility offered by the large number of sources reported by \lat\ is to train 
machine-learning algorithms on populations of known gamma-ray sources to find similar 
candidates in the unassociated sample. \citet{ackermann} characterized the gamma-ray properties of 
pulsars and active galactic nuclei (AGN) in the LAT 11-month catalog \citep{1fgl}, and 
listed unassociated sources 
with similar characteristics. \citet{mirabal} followed a similar approach, finding candidate 
classifications for unassociated 2FGL sources at high Galactic latitudes ($|b| > 10^\circ$), while 
\citet{lee} used a Bayesian approach to find pulsar candidates.

In this work, machine-learning algorithms are used to identify unassociated sources in the 2FGL 
catalog with properties similar to gamma-ray-emitting AGN. Two 
different learning algorithms are trained on the gamma-ray properties of the 
known AGN in the 2FGL catalog. Only the sources selected by both algorithms independently 
are considered AGN candidates, adding robustness to the classification method. In addition, a 
realistic way of estimating the performance of classification methods is presented
that takes into account the differences between the associated and 
unassociated source samples.
Section~\ref{sec:sourceclasses} of this paper describes the properties of the 2FGL catalog. 
Section~\ref{sec:prepare} shows how the data was prepared and which classification algorithms 
were tested, while Section~\ref{sec:optimization} discusses how the algorithms were optimized.
The method for performance estimation 
is discussed in Section~\ref{sec:performance}, and the final results are 
presented in Section~\ref{sec:results}. Finally, Section~\ref{sec:conclusions} summarizes the main 
conclusions of this study.

\section{Source classes in the 2FGL catalog}
\label{sec:sourceclasses}
There are fourteen classes of gamma-ray sources represented in the 2FGL catalog 
(Table~\ref{tab:classes}).
The different types of AGN add up to 60\% of the population. 
The rest of the catalog is distributed among 
unassociated sources (31\%), and source classes with smaller number counts.

In this study, the classification of 
unassociated 2FGL 
sources is approached as a two-class problem, where 
each source is either labeled as ``AGN" or ``non-AGN" (see Table~\ref{tab:classes}). 
Of the total number of associated sources, 
1092 are labeled as AGN and 205 as non-AGN: mostly 
pulsars, 
pulsar wind nebulae
and supernova remnants.

The gamma-ray properties of LAT-detected sources are discussed in detail in \citet{2fgl}. 
Bright AGN exhibit significant flux variability, 
while pulsars show indication of spectral curvature \citep{ackermann}. 
The top panels of Figure~\ref{fig:2fgl_par} show differences between AGN and non-AGN in some 
parameter distributions.
Given these differences, machine-learning algorithms can be trained to recognize 
unassociated sources with AGN-like properties.
Although 
pulsars also have distinct gamma-ray properties, they are not treated as a separate population to 
produce a list of pulsar candidates in this work. Detailed searches for pulsar candidates and multiwavelength 
counterparts 
have received much more attention than AGN in the recent literature
\citep[see, e.g.,][]{keith,kerr,lee,pulsar-cat}.

\section{Data preparation and classification methods}
\label{sec:prepare}
Before starting the learning process, the sample of associated sources (1297 objects) was split 
into two subsamples: training (70\% of the sources) and test (30\%). Subsamples were 
selected using stratified sampling to avoid biasing the parameter distributions.
The training sample was used to train the 
learning algorithms and optimize their performance, while the test sample was set aside to 
evaluate the performance of the classification methods once all the optimizations were made. 

Two quantities characterize the performance of classification algorithms: \rec\ and \far.
The {\it recall} is calculated in this study as the fraction of true AGN that are 
correctly labeled as AGN, and the \far\ is defined as the fraction of non-AGN sources that are 
misclassified as AGN. 

In a first step, a variety of supervised machine-learning classification methods were investigated, 
covering random forest  \citep{rf}, support vector machines 
\citep{svm}, support vector networks \citep{svn}, bayesian classification 
\citep{bayes}, logistic regression \citep{logistic-regression}, nearest-neighbor pattern 
classification \citep{nn-patterns}, and multi-layer perceptrons, also known as neural networks 
\citep{perceptron,mlp}. 
Algorithms were trained using the variables from \citet{ackermann} and default 
settings (e.g., number of iterations). 
The performance parameters were estimated 
using ten-fold cross-validation on the training sample, where the classifier is 
iteratively trained on 90\% of the sample and tested on the remaining 10\%, repeating the 
process ten times until the entire training sample has been tested.

Based on performance, random forest (RF) and neural networks 
(NN) were selected. The choice of two independent algorithms adds robustness to the overall 
classification scheme (RF \& NN), which requires both RF and NN to label a source as AGN for it to 
be considered an AGN candidate.
Combinations of three or more learning algorithms were also explored without showing any 
significant improvement of the performance.

The selection of RF and NN was done after a coarse test over several algorithms. It 
is not excluded that, after a better optimization, other algorithms could slightly improve the 
results presented here.

\section{Optimization of the learning algorithms}
\label{sec:optimization}
Optimization of the RF and NN methods was done by selecting the set of parameters 
that optimizes the learning process, tuning the running parameters of the classification 
algorithms, and adjusting the confidence thresholds to select AGN candidates.

Different attributes from the 2FGL catalog were used during the learning process: \textit{spectral 
index}, $F_i$ (flux in the 5 reported energy bands), \textit{variability}, \textit{curvature}, 
and \textit{significance} (square root of the $TS$ value). 
The best separation power between the populations of AGN and non-AGN  
was found using \textit{spectral index} and seven 
combinations of the abovementioned parameters (many already introduced in 
\citet{ackermann}): $HR_{12}$, 
$HR_{23}$, $HR_{34}$, $HR_{45}$, \textit{hardness slope},  \textit{normalized variability}, and 
\textit{normalized curvature}.
$H R_{ij}$ describes the hardness ratio between the energy fluxes measured in two contiguous 
spectral bands:
\begin{equation}
HR_{ij}=\frac{F_i E_i - F_j E_j}{F_i E_i + F_j E_j}
\end{equation}
where $F_i$ and $E_i$ are, respectively, the flux and mean energy in the $i$-th spectral energy 
band.
A \textit{hardness slope} parameter was also defined as
\begin{equation}
hardness\ slope=HR_{23} - HR_{34}
\end{equation}
which presents a powerful handle to separate possible AGN candidates from pulsar-like sources.
Two additional parameters were also included:
\begin{gather}
normalized\ variability=\frac{variability}{significance} \\
normalized\ curvature=\frac{curvature}{significance}
\end{gather}
Direct use of variables correlated with the overall flux of each source was 
avoided,
and all parameter distributions were renormalized between 0 and 1 to minimize the influence of 
their very different ranges.

  \begin{figure}[tbdp]
 \center
 \includegraphics[width=0.95\columnwidth]{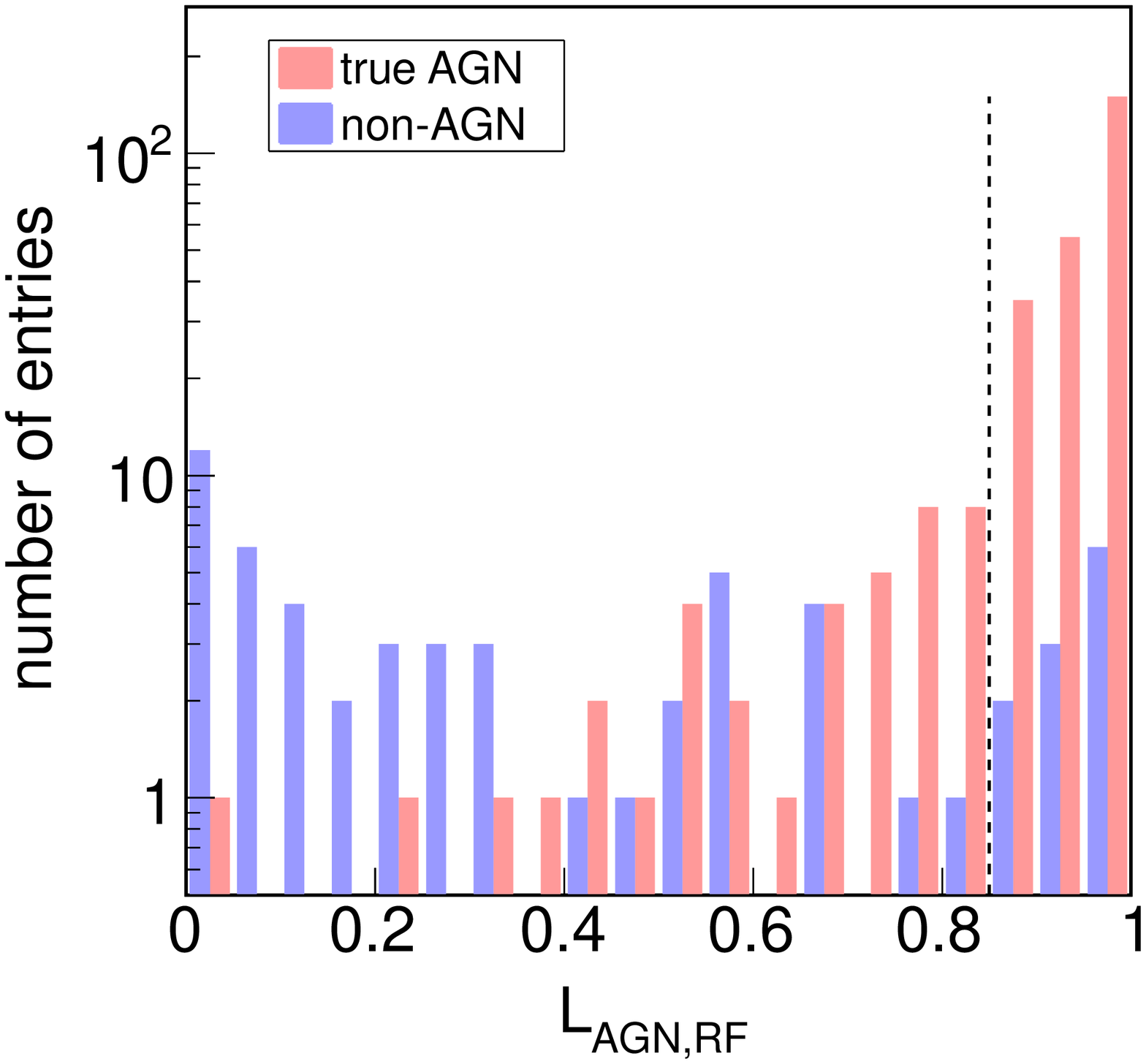} 
 \includegraphics[width=0.95\columnwidth]{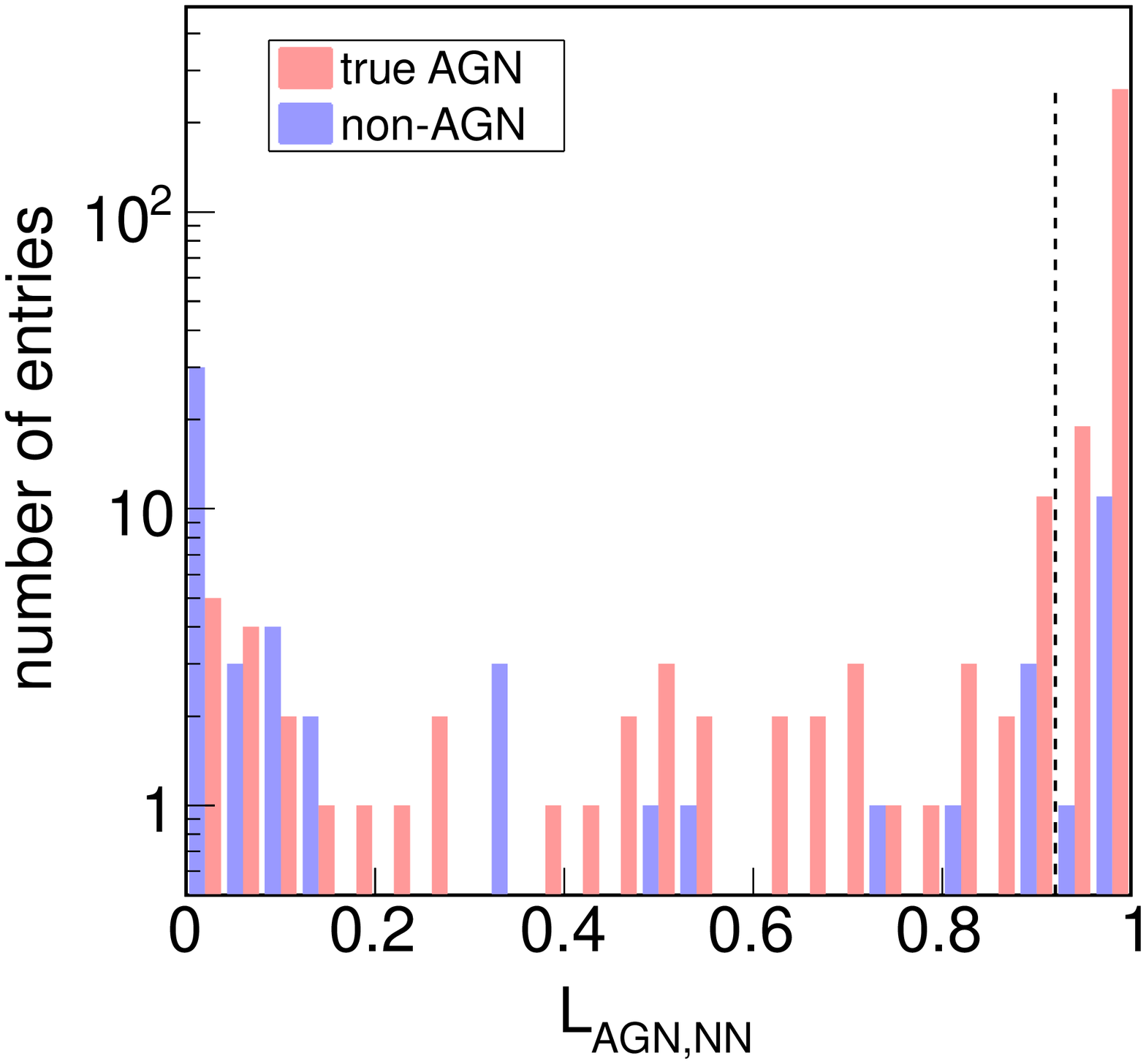} 
 \caption{ Distributions of the likelihood of an AGN classification for ANG and non-AGN sources in the test 
sample. The distributions are shown for random forest (\emph{top 
panel}) and neural networks (\emph{bottom panel}). 
Dashed black lines indicate the likelihood threshold of each algorithm to label a source as AGN.
\label{fig:confidence}}
\end{figure}

\begin{figure*}[tdp]
 \center
 \includegraphics[height=5.5cm]{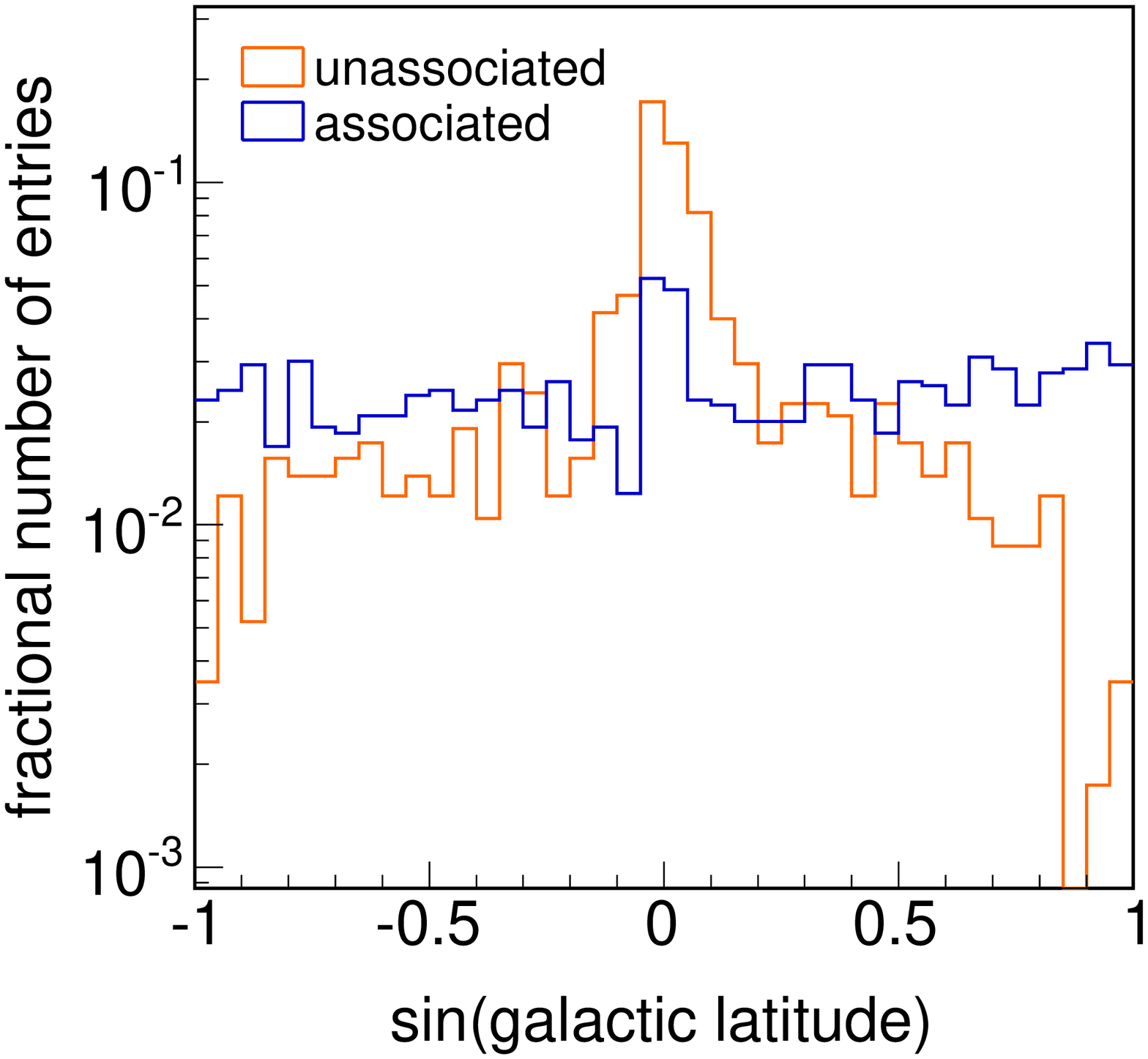} 
 \includegraphics[height=5.5cm]{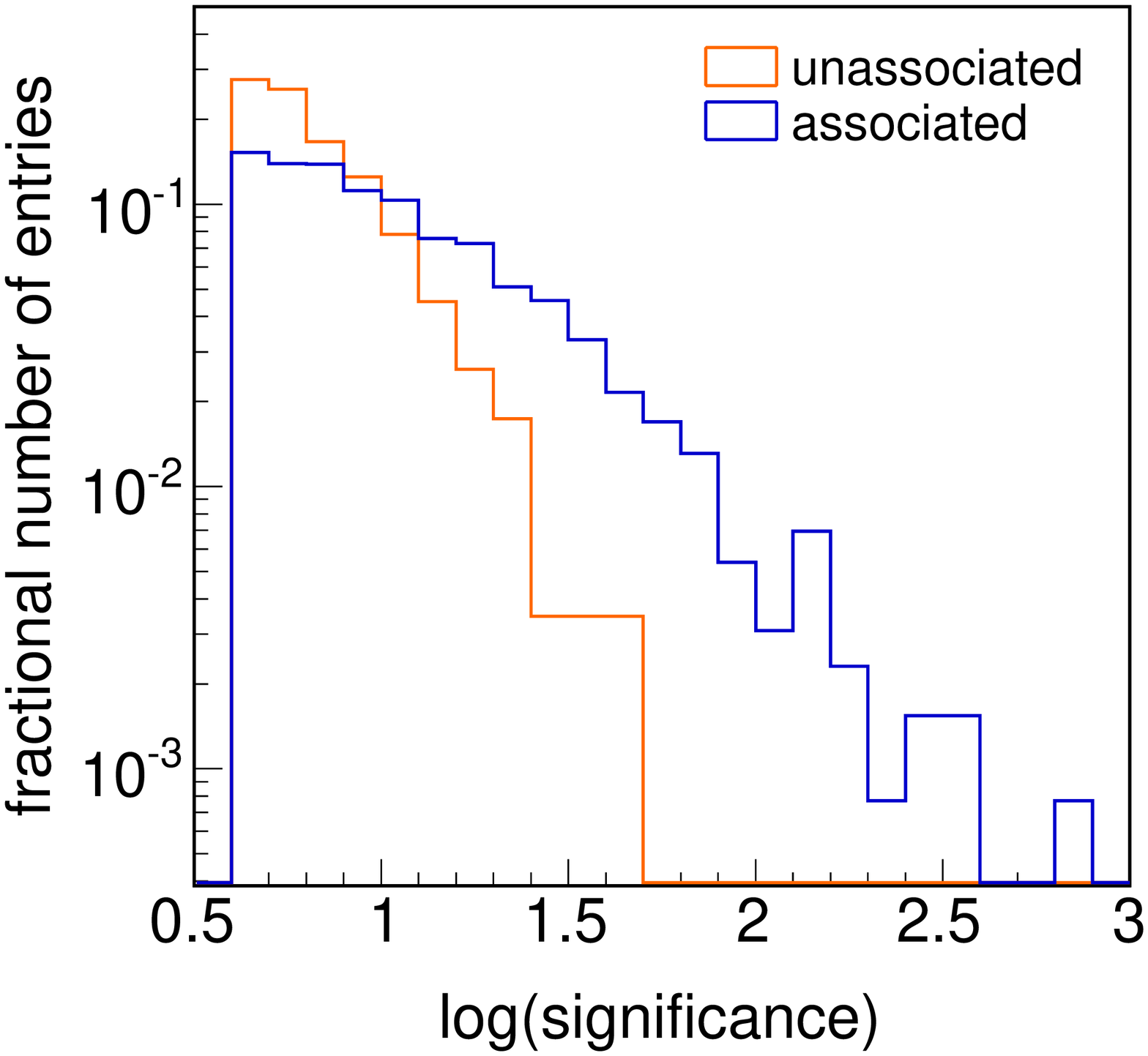}
 \includegraphics[height=5.5cm]{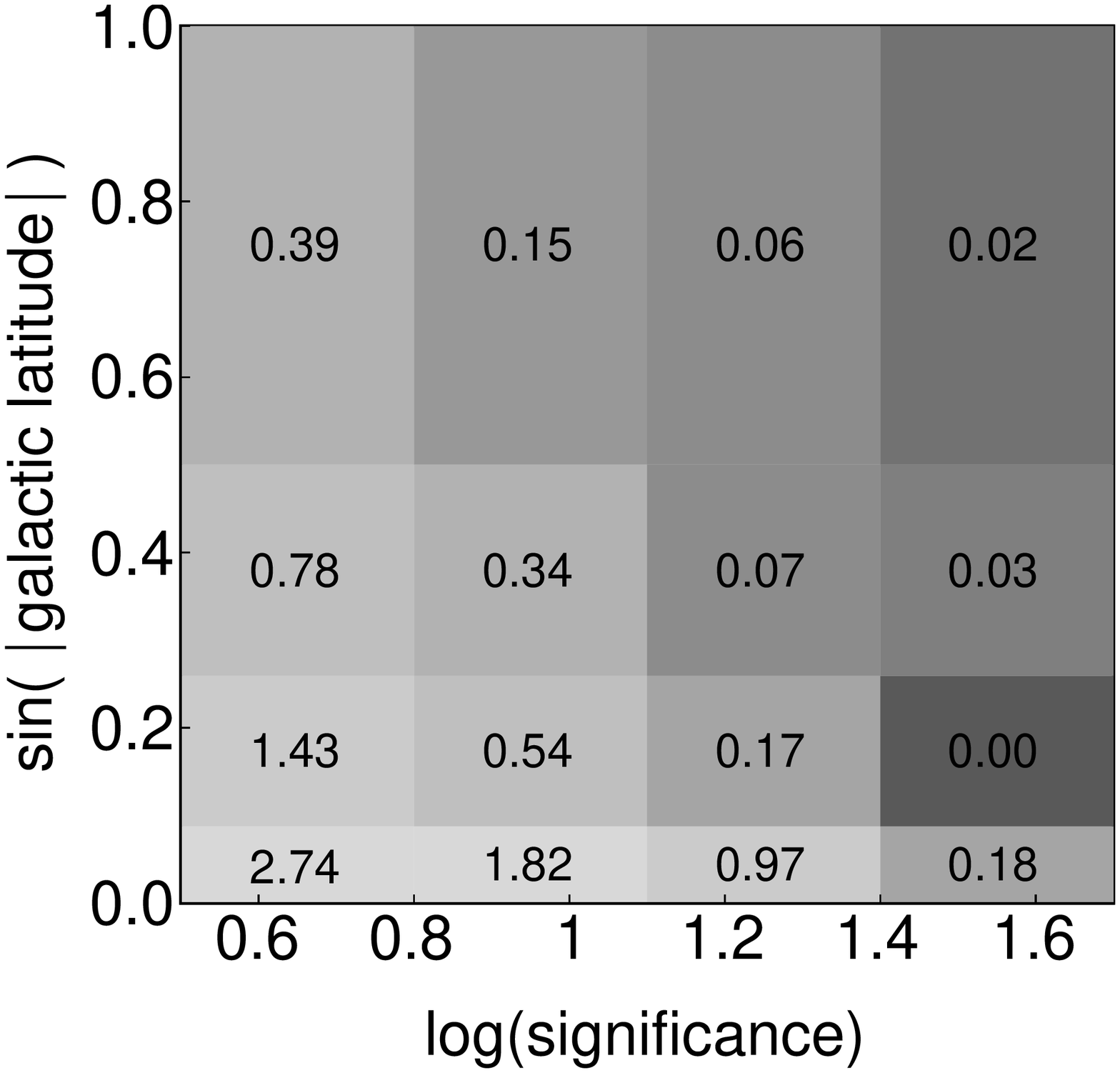}
 \caption{ Galactic latitude (\emph{left panel}) and significance (\emph{middle 
panel}) distributions of associated and unassociated sources in the 2FGL catalog.
\emph{Right panel:} Weights applied to the sources in the test sample to obtain a realistic 
performance estimate, as described in Section~\ref{sec:performance}.
\label{fig:reweighting}}
 \end{figure*}

\begin{table*} [htbp]
\centering
	\begin{tabular}{lcccc}
		\hline\hline
		 & {\it AGN}$\rightarrow$AGN & {\it non-AGN}$\rightarrow$AGN & {\em recall} & 
{\em false-assoc.~rate}\\
		\hline
		random forest    & 289 & 12 & 88.1\% & 16.3\% \\
		neural networks  & 278 & 12 & 84.7\% & 13.5\% \\
		RF \& NN        & 261 &  9 & 79.6\% & 11.2\% \\
		\hline\hline
\end{tabular}
\caption{Performance of the random forest (RF), neural networks (NN), and combined 
algorithm (RF \& NN) evaluated on the test sample, containing 389 sources: 328 AGN and 61 non-AGN. 
Columns show the number of true AGN correctly labeled as AGN, non-AGN misclassified as AGN, and the
recall and false-association rate.}
\label{tab:test}
\end{table*}

 \begin{figure}[htbdp]
\center
\includegraphics[width=0.95\columnwidth]{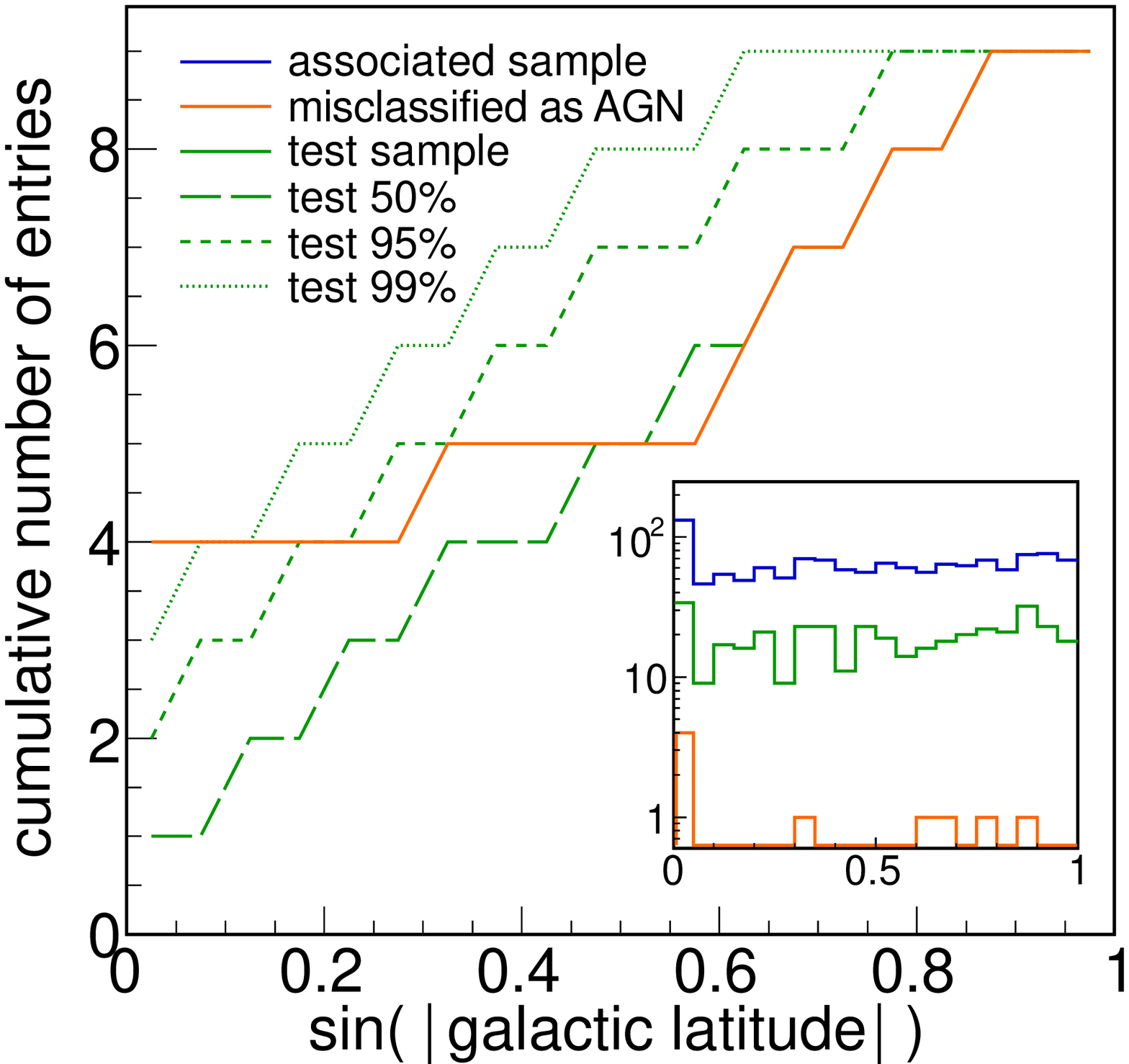} 
\includegraphics[width=0.95\columnwidth]{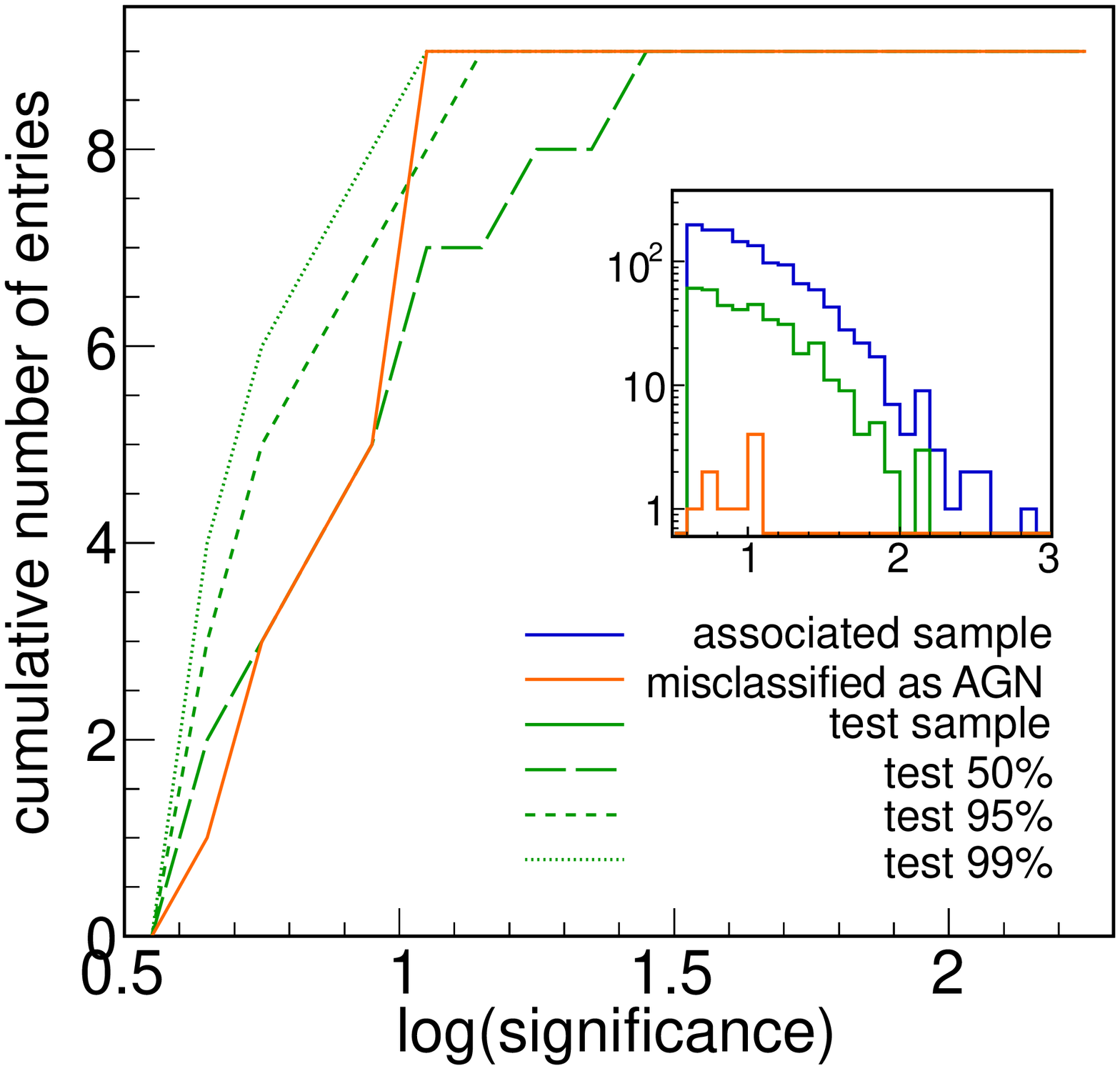} 
\caption{Cumulative distribution of sources in the test sample that were misclassified as AGN 
(orange lines) with increasing Galactic latitude (\emph{top panel}) and significance (\emph{bottom 
panel}).
To test the compatibility of the distributions of misclassified AGN 
with a random sampling of the 389 sources on the test 
sample, 10000 histograms with nine entries randomly 
picked from the distribution of sources in the test sample were generated. 
The long-dashed, dashed, and dotted green lines show the cumulative distributions covering 50\%, 
95\%, and 99\% of the simulated histograms.
Insets show the differential source distributions for 
the associated, test, and misclassified source samples.
 \label{fig:cumulative}}
\end{figure}

The two selected learning algorithms have parameters that can be tuned to improve the 
performance of the method \citep[see][for definitions]{rf,mlp}. 
The RF parameters were adjusted to $number\ of\ trees =100$ 
and $depth\ of\ trees =10$. For NN, values of $number\ of\ cycles=1000$, $learning\ 
rate=0.2$, and $momentum=0.1$ were found to be optimal.

After the learning process, 
RF and NN independently give a likelihood $L_{\mathrm{AGN}}$ of a tested 
source to be an AGN. 
Figure~\ref{fig:confidence} shows likelihood distributions obtained with the RF and 
NN applied to the test sample. Thresholds of $L_{\mathrm{AGN,RF}} \geq 0.85$ and 
$L_{\mathrm{AGN,NN}} \geq 0.92$ were required for each method to label a source as an AGN.
The thresholds were optimized targeting a \far\ of $\sim 10\%$ for the combined 
classification method (RF \& NN).

\section{Performance of the classification method}
\label{sec:performance}

The performance of the classification algorithms is evaluated on the test sample, and is used to 
predict the completeness and number of spurious sources present in the final list of AGN 
candidates.
However, the gamma-ray properties of associated (test) and unassociated sources 
differ in parameters that affect the 
performance of the classification methods. Figure~\ref{fig:reweighting} shows that unassociated 
sources appear to be more clustered at low significances and low Galactic latitudes than 
sources in the associated sample. This is expected, as association probabilities are 
lower for sources with larger location errors, and counterpart catalogs tend to be incomplete near 
the Galactic plane \citep[see][also discussion in Section~\ref{sec:conclusions}]{2fgl}.

Low-significance and low-latitude sources also challenge machine-learning classification 
algorithms. They are often too faint and/or influenced by the bright Galactic foreground to 
extract definitive information on their spectral shape and flux variability. 
As shown in the top panel of Figure~\ref{fig:cumulative}, the 
number of sources incorrectly labeled as AGN at low latitudes has a probability of $<1\%$ of arising
from random sampling the test sample. It is also shown (Figure~\ref{fig:cumulative}, {\it bottom 
panel}) that all misclassified sources have 
$significance < 12$. This scenario has a chance probability of 1\% if 
the performance of the classification algorithm would not depend on significance.
These trends, and the differences between populations shown in Figure~\ref{fig:reweighting}, imply 
that sources in the test sample are easier to classify (or 
less likely to  be misclassified) than unassociated sources. Therefore, 
a \far\ directly evaluated on the test sample will lead to an over-optimistic performance estimate.

To overcome this limitation, sources are binned in significance and Galactic latitude. Then, 
weights are calculated as $w_{ij} = N^\mathrm{ua}_{ij}/N^\mathrm{a}_{ij}$, where 
$N^\mathrm{a}_{ij}$ and $N^\mathrm{ua}_{ij}$ are, respectively, the number of associated and 
unassociated sources in the $i,j$-th bin. The actual binning and weight values are shown in the 
right panel of Figure~\ref{fig:reweighting}. The \far\ is then estimated as
\begin{equation}
\mathit{false\mhyphen association\ rate} 
  = \frac{\sum\limits_{i,j} N_{ij}^\mathrm{fa} \cdot w_{ij}} 
{\sum\limits_{i,j} N_{ij}^\mathrm{AGN} \cdot w_{ij}}
\end{equation}
where $N^\mathrm{fa}_{ij}$ is the number of sources misclassified as AGN on each bin, and 
$N^\mathrm{AGN}_{ij}$ the number of sources labeled as AGN. The use of weights
corrects the bias introduced by the differences between source populations, giving a 
realistic estimate of the \far.

The performance of the classification algorithm (RF \& NN), 
together with the individual performance of each learning method, is shown in Table~\ref{tab:test}.
The  algorithm is expected to recognize 80\% of the AGN present in the unassociated sample, with a 
\far\ of 11\%. 

 \begin{figure}[htpb]
 \centering
 \includegraphics[width=0.95\columnwidth]{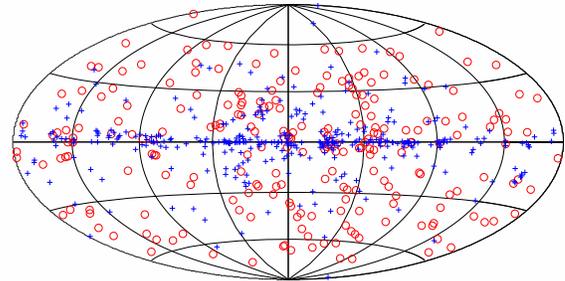} 
 \caption{Sky distribution in Galactic coordinates of all unassociated 2FGL sources. AGN 
candidates are shown as red circles, while blue crosses indicate 
sources that were not labeled as AGN. Adapted from \citet{icrc}.  \label{fig:skymap}}
 \end{figure}

\section{List of AGN candidates}
\label{sec:results}
The classification algorithm (RF \& NN) was applied to the sample of unassociated 2FGL 
sources to 
produce the list of high-confidence AGN candidates shown in Table~\ref{tab:sourcelist}, which is 
the main result of this study. A total of 231 AGN candidates are found among the 576 
unassociated sources that were studied.
The sky distribution of the AGN candidates, together 
with the unassociated sources that were not conclusively labeled, is shown in 
Figure~\ref{fig:skymap}. 

The 231 AGN candidates constitute 40\% of the 2FGL unassociated sources. This is in line 
with estimates from similar works, that predicted $35\%-55\%$ of unassociated gamma-ray 
sources detected by \lat\ to be AGN \citep{ackermann,ferrara}. 
Out of 159 AGN candidates at $|b|\geq10^\circ$, 156 are also listed 
as likely AGN according to a previous work by \citet{mirabal} that focused only 
on 2FGL sources outside the galactic plane.

Sources potentially confused with interstellar emission (flagged with a ``c'' designator in the 
2FGL catalog) were treated as regular sources throughout the analysis. Those constitute 22\% of the 
unassociated sample, and have typically low detection significances. The list of 231 AGN candidates 
contains 22 sources with ``c'' designator (9\%), showing that confused sources were less 
likely to be labeled as high-confidence AGN, as expected for weak sources where the spectral and 
variability properties are less certain.

The classification algorithm finds 11 sources at $|b|\geq10^\circ$ with no significant similarities 
with known AGN ($L_{\mathrm{AGN,RF}} < 0.5$ \& $L_{\mathrm{AGN,NN}} < 0.5$). These could 
potentially be interesting, as searches for dark matter annihilation or decay signals from dark 
subhaloes target high-latitude unassociated sources with no obvious counterparts  
\citep[][]{nieto,zechlin}. However, all but \object[2FGL J0538.5-0534c]{2FGL~J0538.5-0534c} are 
pulsar candidates \citep{ackermann,mirabal,lee} or have known X-ray counterparts \citep{suzaku}.

\begin{table} [b]
\centering
	\begin{tabular}{lcccccc}
		\hline\hline
		 2FGL name &	R.A. [$^\circ$]& decl. [$^\circ$]& $L_\mathrm{RF}$ & 
$L_\mathrm{NN}$ & (1) & (2) \\
		\hline
		J0004.2+2208	&1.056	&22.137	& 0.98  & 0.97 & A &  \\
		J0014.3-0509	&3.581	&-5.153	& 1.00  & 1.00 & A &  \\
		J0031.0+0724	&7.775	&7.414	& 0.99  & 1.00 & A & b \\
		J0032.7-5521	&8.179	&-55.356& 1.00  & 1.00 & A &  \\
		J0048.8-6347	&12.218	&-63.79	& 0.90  & 0.92 & A & b \\
		J0102.2+0943	&15.553	&9.726	& 1.00  & 1.00 & A &  b\\
		J0103.8+1324    &15.953 &13.401	& 0.96  & 1.00 & A &  b\\
		J0116.6-6153	&19.174 &-61.887& 1.00  & 1.00 & A &  ab\\
		J0133.4-4408	&23.364 &-44.142& 0.99  & 1.00 & A &  ab\\
		J0143.6-5844	&25.917 &-58.745& 0.98  & 1.00 & A &  abc\\
		\hline\hline
\end{tabular}
\caption{List of high-confidence AGN candidates, ordered by R.A.
(This table is available in its entirety in 
a machine-readable form in the online journal. A portion is shown here for guidance regarding its 
form and content.)
(1) Class predicted by \citet{mirabal}; A:AGN, -:Uncertain.
(2) Counterparts: \textup{a}: infrared counterpart in \citet{massaro-ir}, \textup{b}: X-ray counterpart in \citet{paggi}, \textup{c}: AGN candidate in \citet{acero}. 
\label{tab:sourcelist}}
\end{table}

\section{Discussion and conclusions}
\label{sec:conclusions}

This work studied the sample of unassociated gamma-ray sources in the \lat\ 2FGL catalog, finding 
231 AGN candidates based on their gamma-ray properties. 
Two independent machine-learning algorithms (random forest and neural networks) were used to asses 
the likelihood of each source to be an AGN, and intersected to
add robustness to the 
classification method and reduce the number of false associations.

The study includes for the first time an estimate of the \far\ that takes into 
account the differences between associated and unassociated gamma-ray sources. 
By evaluating the performance using a test sample weighted in significance and Galactic latitude, 
the obtained  11\% \far\ can be considered a realistic estimate of the fraction of spurious 
sources present in the AGN-candidate list. 
\citet{ackermann} obtained a lower \far\ directly evaluated on the test sample,
which is likely an optimistic performance estimate as discussed in Section~\ref{sec:performance}.
Similarly, \citet{mirabal}
used cross-validation on the training sample, which is 
known to give an optimistic performance, as the same sources used to optimize the 
classification method are used to calculate the \far.

\begin{figure}[tdb]
\center
\includegraphics[width=0.95\columnwidth]{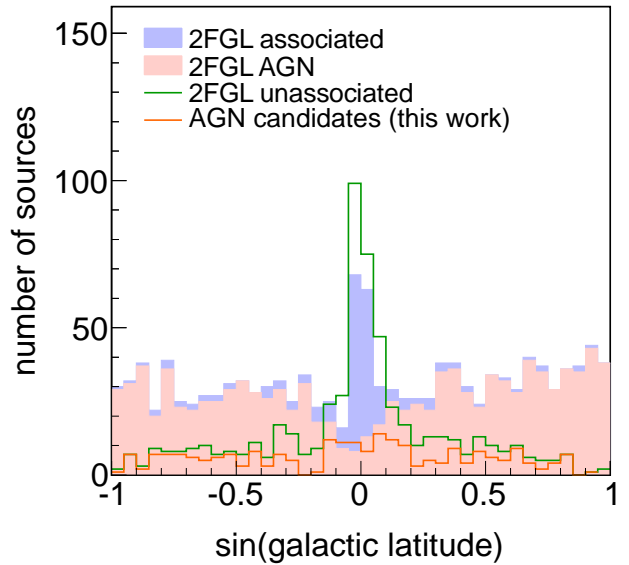} 
\caption{Galactic latitude distribution of the 2FGL sources. Shown separately are associated and 
unassociated sources, known AGN, and AGN candidates identified in this work.
 \label{fig:lat}}
\end{figure}

The list of AGN candidates (Table~\ref{tab:sourcelist}) covers the whole sky, studying for the 
first time the strip that covers the Galactic plane, where 
more than 50\% of the unassociated 2FGL sources are located. 
About 210 AGN are expected at $|b|\leq10^\circ$ extrapolating from high-latitude observations 
\citep{2lac}, while only 104 are listed in the 2FGL catalog.
Even though low-latitude sources are 
harder to classify, 
72 AGN candidates were found at $|b|\leq10^\circ$ (see Figure~\ref{fig:lat}), which could be a 
significant fraction of the missing AGN close to the Galactic plane.
At $|b|\geq10^\circ$, the list of AGN candidates 
 is in good agreement with previous work by  
\citet{mirabal}. Their study found 60 additional AGN candidates, which could be a 
combination of their method being more sensitive, as it was trained on a 
cleaner sample of high-latitude sources, and a lower confidence threshold to identify AGN 
candidates.

Close to the Galactic plane, AGN are difficult to identify due to optical 
extinction and the bright foreground in radio and soft X-rays. Counterpart 
catalogs are usually incomplete at low latitudes or skip the Galactic plane altogether 
\citep[e.g.,][]{cgrabs,bzcat}. 
Galactic absorption for gamma rays  is negligible below 10\,TeV \citep{moskalenko}, making
low-latitude AGN detectable in the gamma-ray band but difficult to catalog at lower frequencies.
In fact, numerous identifications of AGN  behind the Galactic plane have been triggered by 
gamma-ray detections at GeV \citep{muk-2016,mirabal-halpern,kara} and TeV energies 
\citep{hess-1943,0521}.

The list of 231 candidate AGN presented here cannot be considered source associations, but only 
objects likely to be associated with AGN.
In case of gamma-ray-emitting AGN, detectable levels of non-thermal emission in radio, optical, and 
X-ray frequencies are expected, and follow-up observations in those bands are 
needed to unambiguously identify the nature of the gamma-ray emission. 
Observations in the X-ray band (0.2-10\,keV) have been successful in finding counterparts of 
unidentified gamma-ray sources \citep[e.g.,][]{muk-2016}.
The angular resolution of X-ray telescopes ($\sim 0.3^\prime$ for {\it Swift}-XRT) can resolve 
individual sources inside the typical $\sim 5^\prime$ error box of 
unassociated 2FGL sources. Follow-up observations in radio and 
optical spectroscopy of candidate X-ray counterparts can then provide a solid AGN identification 
and spectral class 
\citep[e.g.,][]{halpern-2016}. 

The {\it Swift} X-ray satellite has observed a good fraction of the 2FGL unassociated sources. 
A complete summary of these observations can be found in 
\url{http://www.swift.psu.edu/unassociated/} \citep[][]{stroh}. 
So  far, 135 out of the 231 candidate AGN have  
at least 2\,ks of {\it Swift}-XRT exposure. 
A good fraction of these have also been analyzed 
in \citet{paggi}, finding 85 sources with at least one point-like X-ray counterpart. 
Infrared counterparts with AGN-like spectra have  been identified for 56 of the AGN candidates 
in the WISE all-sky survey \citep{massaro-ir}. Recently, \citet{acero} presented a multiwavelength 
study of seven unassociated sources where four AGN candidates were investigated and confirmed to 
have AGN-like properties in the radio and X-ray bands. These counterparts are listed in Table~\ref{tab:sourcelist}.

Counterpart catalogs of gamma-ray-emitting AGN candidates are mostly based on the population 
of AGN detected by EGRET \citep{egret-agn}, that contained a large number of flat-spectrum 
radio quasars (FSRQ, 71\%) and fewer BL~Lac-type objects (27\%).
Similar FSRQ/BL~Lac ratios are found in counterpart catalogs such as {\em CGRaBS} 
\citep[84\% FSRQ\,/\,10\% BL~Lac,][]{cgrabs} or {\em BZCAT} \citep[54\%/39\%,][]{bzcat}. However, 
BL~Lacs are more numerous than FSRQ in the 2FGL catalog (34\%/40\%, see Table~\ref{tab:classes}). 
LAT-detected BL~Lacs
have a median radio flux density of 86\,mJy \citep{2lac}, with a 
low-flux tail extending well below the completeness limit of the {\em CRATES/CGRaBS} catalog 
\citep[65\,mJy,][]{crates}. 
The potential deficit of BL~Lacs in counterpart catalogs suggests that a fraction of unassociated 
2FGL sources might indeed be BL~Lac-type blazars that have not yet been cataloged. 
This could become relevant in searches for TeV-emitting AGN with present ground-based observatories 
\citep[e.g.,][]{massaro-tev}, and prospects for future installations like CTA \citep{cta-agn}, as 
the harder gamma-ray spectra of BL~Lacs favor their detection at TeV energies over FSRQ.

Follow-up studies on the AGN candidates presented here will reduce the number of 
unassociated gamma-ray sources and yield a more complete picture of the characteristics of 
gamma-ray-loud AGN. Additionally, future observations could prove whether the population of 
gamma-ray-emitting BL~Lacs extends to sources with low radio flux density. If confirmed, gamma-ray 
emission from BL~Lacs with luminosities $\lesssim 10^{44}\,\mathrm{erg}\,\mathrm{s}^{-1}$ will give 
additional information on the low end of BL~Lac luminosity function in the gamma-ray band 
\citep{ajello}, which is a key ingredient to estimate their contribution to the isotropic diffuse 
gamma-ray background \citep{background}.

\acknowledgments

\begin{small}
This work was supported by DFG (SFB\,823, SFB\,876), DAAD (PPP USA) and HAP. 
ME acknowledges support from NASA grant NNX12AJ30G. The authors thank Reshmi Mukherjee and 
Daniel Nieto for very useful discussions on the structure of the paper and the 
interpretation of the results, and Sabrina Einecke and Ann-Kristin Overkemping for proof-reading 
the manuscript.
This study used RapidMiner \citep{Mierswa2006}: an open-source data mining framework developed at 
Technische Universit\"{a}t Dortmund and maintained and distributed by Rapid-I.
\end{small}


\begin{thebibliography}{99}


\bibitem[Abdo et al.(2010a)]{background} Abdo, A.~A., Ackermann, M., Ajello, M., et al.\ 2010, Phys. 
Rev. Lett., 104, 101101 

\bibitem[Abdo et al.(2010b)]{1fgl} Abdo, A.~A., Ackermann, M., Ajello, M., et al.\ 2010, \apjs, 188, 
405 

\bibitem[Abdo et al.(2013)]{pulsar-cat} Abdo, A.~A., Ajello, M., Allafort, A., et al.\ 2013, \apjs, 
208, 17 

\bibitem[Abramowski et al.(2011)]{hess-1943} Abramowski, A., Acero, F., Aharonian, F., et al.\ 
2011, \aap, 529, A49 

\bibitem[Acero et al.(2013)]{acero} Acero, F., Donato, D., Ojha, R., et al.\ 2013, \apj, 779, 133 

\bibitem[Ackermann et al.(2011)]{2lac} Ackermann, M., Ajello, M., Allafort, A., et al.\ 2011, \apj, 
743, 171 

\bibitem[Ackermann et al.(2012)]{ackermann} Ackermann, M., Ajello, M., Allafort, A., et al.\ 2012, 
\apj, 753, 83 

\bibitem[Ajello et al.(2013)]{ajello} Ajello, M., Romani, R.~W., Gasparrini, D., et al.\ 2013, \apj, 780, 73

\bibitem[Archambault et al.(2013)]{0521} Archambault, S., Arlen, T., Aune, T., et al.\ 2013, \apj, 
776, 69 

\bibitem[Berger(1980)]{bayes} Berger, J.~O.\ 1980, ``Statistical decision theory and bayesian 
analysis'' (Springer, New York)

\bibitem[Bertsekas \& Tsitsiklis(1995)]{Bertsekas:1995jl} Bertsekas, D.~P. \& Tsitsiklis, 
J.~N.\ 1995, Neuro-dynamic programming: an overview Vol. 1, 560  

\bibitem[Breiman(2001)]{rf} Breiman, L.\ 2001, Machine Learning, Vol. 45

\bibitem[Chang \& Lin(2011)]{svm} Chang, C.-C. \& Lin, C.-J.\ 2011, ACM Transactions on Intelligent 
Systems and Technology, 2, 27 

\bibitem[Cortes \& Vapnik(1995)]{svn} Cortes, C. \& Vapnik, V.\ 1995, Machine Learning, 20, 273 

\bibitem[Cover \& Hart(1967)]{nn-patterns} Cover, T. \& Hart, P.\ 1967, IEEE Transactions on 
Information Theory, 13, 21

\bibitem[Crawford et al.(2006)]{crawford} Crawford, F., Roberts, 
M.~S.~E., Hessels, J.~W.~T., et al.\ 2006, \apj, 652, 1499 

\bibitem[Cybenko(1989)]{mlp} Cybenko, G.\ 1989, Math. Control Signals Systems, 2, 303

\bibitem[Doert \& Errando(2013)]{icrc} Doert, M., \& Errando, M.\ 2013, to appear in Proc. 33rd Int. 
Cosmic Ray Conf. (Rio de Janeiro), arXiv:1306.6529 

\bibitem[Ferrara et al.(2012)]{ferrara} Ferrara, E.~C., Ojha, 
R., Monzani, M.~E., \& Omodei, N.\ 2012, in Proc. 2012 Fermi \& Jansky - eConf 
C1111101 (arXiv:1206.2571) 

\bibitem[Gehrels et al.(2000)]{gehrels} Gehrels, N., Macomb, 
D.~J., Bertsch, D.~L., Thompson, D.~J., 
\& Hartman, R.~C.\ 2000, \nat, 404, 363 

\bibitem[Ghirlanda et al.(2010)]{ghirlanda} Ghirlanda, G., 
Ghisellini, G., Tavecchio, F., \& Foschini, L.\ 2010, \mnras, 407, 791 

\bibitem[Halpern et al.(2001)]{halpern-2016} Halpern, J.~P., 
Eracleous, M., Mukherjee, R., \& Gotthelf, E.~V.\ 2001, \apj, 551, 1016 

\bibitem[Hartman et al.(1999)]{3eg} Hartman, R.~C., 
Bertsch, D.~L., Bloom, S.~D., et al.\ 1999, \apjs, 123, 79 

\bibitem[Hartman et al.(1979)]{sas2} Hartman, R.~C., 
Kniffen, D.~A., Thompson, D.~J., et al.\ 1979, \apj, 230, 597 

\bibitem[Healey et al.(2008)]{cgrabs} Healey, S.~E., Romani, 
R.~W., Cotter, G., et al.\ 2008, \apjs, 175, 97 

\bibitem[Healey et al.(2007)]{crates} Healey, S.~E., Romani, 
R.~W., Taylor, G.~B., et al.\ 2007, \apjs, 171, 61 

\bibitem[Hosmer \& Lemeshow(2000)]{logistic-regression} Hosmer, D., \& Lemeshow, S. 2000, ``Applied 
Logistic Regression, 2nd edn.'' (John Wiley \& Sons, Inc., New York)

\bibitem[Kaaret \& Cottam(1996)]{kaaret} Kaaret, P., \& Cottam, J.\ 1996, \apjl, 462, L35 

\bibitem[Kara et al.(2012)]{kara} Kara, E., Errando, M., 
Max-Moerbeck, W., et al.\ 2012, \apj, 746, 159 

\bibitem[Keith et al.(2011)]{keith} Keith, M.~J., Johnston, 
S., Ray, P.~S., et al.\ 2011, \mnras, 414, 1292 

\bibitem[Kerr et al.(2012)]{kerr} Kerr, M., Camilo, F., 
Johnson, T.~J., et al.\ 2012, \apjl, 748, L2 

\bibitem[Lamb \& Macomb(1997)]{lamb} Lamb, R.~C., \& Macomb, D.~J.\ 1997, \apj, 488, 872 

\bibitem[Lee et al.(2012)]{lee} Lee, K.~J., Guillemot, L., 
Yue, Y.~L., Kramer, M., \& Champion, D.~J.\ 2012, \mnras, 424, 2832 

\bibitem[Massaro et al.(2009)]{bzcat} Massaro, E., Giommi, P., Leto, C., et al.\ 2009, \aap, 495, 
691 

\bibitem[Massaro et al.(2013a)]{massaro-radio} Massaro, F., D'Abrusco, 
R., Giroletti, M., et al.\ 2013a, \apjs, 207, 4 

\bibitem[Massaro et al.(2013b)]{massaro-ir} Massaro, F., D'Abrusco, 
R., Paggi, A., et al.\ 2013b, \apjs, 206, 13 

\bibitem[Massaro et al.(2013c)]{massaro-tev} Massaro, F., Paggi, A., 
Errando, M., et al.\ 2013c, \apjs, 207, 16

\bibitem[Merck et 
al.(1996)]{merck} Merck, M., Bertsch, D.~L., Dingus, B.~L., et al.\ 1996, \aaps, 120, 
465 

\bibitem[Mierswa(2006)]{Mierswa2006} Mierswa, I. et al.\ 2006, in Proc. 12th ACM SIGKDD 
Int. Conf. on Knowledge discovery and data mining, 935 

\bibitem[Mirabal \& Halpern(2009)]{mirabal-halpern} Mirabal, N., \& Halpern, J.~P.\ 2009, \apjl, 
701, L129 

\bibitem[Mirabal et al.(2012)]{mirabal} Mirabal, N., Fr{\'{\i}}as-Martinez, V., Hassan, T., 
and Fr{\'{\i}}as-Martinez, E.\ 2012, \mnras, 424, L64 

\bibitem[Moskalenko et al.(2006)]{moskalenko} Moskalenko, I.~V., 
Porter, T.~A., \& Strong, A.~W.\ 2006, \apjl, 640, L155 

\bibitem[Mukherjee et al.(1997)]{egret-agn} Mukherjee, R., 
Bertsch, D.~L., Bloom, S.~D., et al.\ 1997, \apj, 490, 116 

\bibitem[Mukherjee et al.(1995)]{mukherjee} Mukherjee, R., 
Bertsch, D.~L., Dingus, B.~L., et al.\ 1995, \apjl, 441, L61 

\bibitem[Mukherjee et al.(2000)]{muk-2016} Mukherjee, R., 
Gotthelf, E.~V., Halpern, J., \& Tavani, M.\ 2000, \apj, 542, 740 

\bibitem[Nieto et al.(2011)]{nieto} Nieto, D., Aleksi{\'c}, 
J., Barrio, J.~A., et al.\ 2011, in Proc. 33rd Int. 
Cosmic Ray Conf. (Beijing) arXiv:1109.5935

\bibitem[Nolan et al.(2012)]{2fgl} Nolan, P.~L., Abdo, 
A.~A., Ackermann, M., et al.\ 2012, \apjs, 199, 31 

\bibitem[Ozel et 
al.(1988)]{ozel} Ozel, M.~E., Schlickeiser, R., Sieber, W., \& Younis, S.\ 1988, 
\aap, 200, 195 

\bibitem[Paggi et al.(2013)]{paggi} Paggi, A., Massaro, F., 
D'Abrusco, R., et al.\ 2013, \apjs, 209, 9

\bibitem[Romero et 
al.(1999)]{romero} Romero, G.~E., Benaglia, P., \& Torres, D.~F.\ 1999, \aap, 348, 868 

\bibitem[Rosenblatt(1961)]{perceptron} Rosenblatt, F.\ 1961, ``Principles of Neurodynamics: 
Perceptrons and the Theory of Brain Mechanisms'' (Spartan Press, Washington)

\bibitem[Sol et al.(2013)]{cta-agn} Sol, H., Zech, A., Boisson, 
C., et al.\ 2013, Astroparticle Physics, 43, 215 

\bibitem[Sowards-Emmerd et al.(2003)]{sowards-emmerd} Sowards-Emmerd, 
D., Romani, R.~W., \& Michelson, P.~F.\ 2003, \apj, 590, 109 

\bibitem[Stroh 
\& Falcone(2013)]{stroh} Stroh, M.~C., \& Falcone, A.~D.\ 2013, \apjs, 207, 28 

\bibitem[Swanenburg et al.(1981)]{cosb} Swanenburg, B.~N., Bennett, K., Bignami, G.~F., et al.\ 
1981, \apjl, 243, L69 

\bibitem[Takeuchi et al.(2013)]{suzaku} Takeuchi, Y., Kataoka, J., Maeda, K., et al.\ 2013, \apjs, 
208, 25 

\bibitem[Zechlin \& Horns(2012)]{zechlin} Zechlin, H.-S., \& Horns, D.\ 2012, J. Cosmol. Astropart. 
Phys., 11, 50 

\end{thebibliography}
\end{document}